\documentclass[sor, jor, reprint]{revtex4-2}
\def\abstractname{ }

\usepackage[utf8]{inputenc}
\usepackage{amsmath, amssymb, commath, mathtools}
\usepackage{graphicx}
\usepackage{standalone}
\usepackage{tikz}
\usetikzlibrary{calc}
\usepackage{color}
\usepackage{times}
\usepackage{booktabs}
\PassOptionsToPackage{hyphens}{url}
\usepackage{placeins}
\usepackage{natbib}

\graphicspath{{fig/}}

\newtheorem{assumption}{Assumption}

\begin{document}
	
{\large\begin{center}Preprint submitted to Annual Reviews in Control: Special Issue on Covid-19 \end{center}}

\title{Smart Testing and Selective Quarantine \\ for the Control of Epidemics}

\author{\vspace{10pt}Matthias~Pezzutto}
\affiliation{\hbox{Department of Information Engineering, University of Padova (Italy)}\\ \hbox{email: matthias.pezzutto@phd.unipd.it, schenato@dei.unipd.it}}

\author{Nicolas~Bono~Rossello}
\affiliation{\hbox{Service d’Automatique et d’Analyse des Systèmes, Université Libre de Bruxelles (Belgium)} \\ \hbox{e-mail: \{nbonoros,egarone\}@ulb.ac.be}}

\author{Luca~Schenato}
\affiliation{\hbox{Department of Information Engineering, University of Padova (Italy)}\\ \hbox{email: matthias.pezzutto@phd.unipd.it, schenato@dei.unipd.it}}

\author{Emanuele~Garone\vspace*{10pt}}
\affiliation{\hbox{Service d’Automatique et d’Analyse des Systèmes, Université Libre de Bruxelles (Belgium)} \\ \hbox{e-mail: \{nbonoros,egarone\}@ulb.ac.be}}

\date{22 December 2020}

\begin{abstract}
This paper is based on the observation that, during Covid-19 epidemic, the choice of which individuals should be tested has an important impact on the effectiveness of selective confinement measures. This decision problem is closely related to the problem of optimal sensor selection, which is a very active research subject in control engineering.
The goal of this paper is to propose a policy to smartly select the individuals to be tested. The main idea is to model the epidemics as a stochastic dynamic system and to select the individual to be tested accordingly to some optimality criteria, e.g. to minimize the probability of undetected asymptomatic cases. Every day, the probability of infection of the different individuals is updated making use of the stochastic model of the phenomenon and of the information collected in the previous days.
Simulations for a closed community of 10’000 individuals show that the proposed technique, coupled with a selective confinement policy, can reduce the spread of the disease while limiting the number of individuals confined if compared to the simple contact tracing of positive and to an off-line test selection strategy based on the number of contacts.
\end{abstract}

\vspace*{5pt}

\maketitle

\section{Introduction}

During the Covid-19 epidemic, one of the limiting factors that affected the capability to handle the spread of the disease was the limited number of available tests. This lack of information has created major issues in several countries and promoted the idea that testing is essential in the control of an epidemic~(\cite{salath_covid19_2020}).

Recent research works support the importance of testing to effectively control an epidemic, see~\cite{brotherhood_economic_2020, wang_analytical_2020, eichenbaum_macroeconomics_2020}. In this regard, the selection of the individuals to be tested has become a major concern in many countries. However, to the best of the authors' knowledge, research on how to define these testing policies is still at a very early stage~(\cite{nowzari_analysis_nodate}).

This observation is testified by the {\it de facto} policies applied by decision makers during the Covid-19 epidemic. Among the various policies we can mention the use of contact tracing of individuals exposed to positive cases~(\cite{cereda_early_2020}), contact tracing combined with additional random testing~(\cite{shim_transmission_2020}), the use of exhaustive control of new arrivals in isolated communities~(\cite{wang_response_2020}), and the testing of people with high number of human interaction such as health care personnel (\cite{padula_why_2020}). It is worth to note that most of these strategies rely  on the appearance of symptomatic cases and required the use of hard lockdown policies to be effective.

Interestingly enough, the selection of individuals to test has important similarities with the problem of sensor selection for state estimation in the context of Wireless Sensor Networks. In both cases only a limited amount of information on a partially unknown process can be retrieved due to a limited amount of resources, i.e. the number of available tests or the channel bandwidth, respectively. The objective is to optimize where to collect the measurements based on the available information and on the model of the process.
Sensor selection has been an active field in the last two decades: a method based on convex optimization is proposed by \cite{joshi2008sensor}, a stochastic policy is studied by \cite{gupta2006stochastic} and the optimal periodic policy for two sensors is given by \cite{shi2011optimal}. In the case of a general number of sensors the problem has been explored by \cite{vitus2012efficient} over a finite horizon, by \cite{mo2014infinite} over the infinite horizon, and for a general number of independent dynamical systems by \cite{han2017optimal}. However most of available works on sensor selection focus on real-valued dynamical systems, while the case where  the process state assumes values from a finite set is at the best of our knowledge still largely unexplored. 

The first step to propose an effective smart testing is the selection of an adequate model to monitor the epidemic. Compartmental epidemic models proved to provide accurate estimations of the dynamics of an epidemic~(\cite{brauer_compartmental_2008}). These models can be divided in deterministic models, governed by differential equations~(\cite{mccluskey_complete_2010}), or stochastic models, where the heterogeneity of small communities can be better represented~(\cite{bjornstad_dynamics_2002,lopez-herrero_cumulative_2017}). New models tailored for the Covid-19 case have been developed seeking for more suitable approaches for the design of control strategies, e.g.~\cite{giordano_modelling_2020, casella2020canCOVID, franco2020feedbackSIR}. However, the nature of compartmental models implies an homogeneously distributed population with random mixing between individuals, which does not inform about the granular distribution of the disease. To model the granularity of the spread of a disease, network diffusion models provide a better insight of the population's distribution and allow to identify the critical clusters of the spreading.

The most common network diffusion models are based on  Stochastic Cellular Automata (SCA), where the spread of the disease depends on the interaction between neighboring cells~(\cite{mikler_modeling_2005,white_modeling_2007}). This idea has been lately extended to more complex network topologies~(\cite{li_analysis_2014,keeling_networks_2005}). In these complex network models the interactions between individuals are modelled as the edges of a graph. This representation makes it possible to also model time-varying interactions, as well as selective quarantine policies (e.g. by removing the connections of certain individual with the rest of the population). From the theoretical viewpoint it is possible to prove that any SCA model is equivalent to a Markov chain (\cite{ruhi_sirs_2015}). As we will discuss later on in this paper, this fact, although important from the theoretical viewpoint, is however not very useful in practice as the resulting  Markov chain has a number of states that is exponential in the number of the states of SCA. 

It is important to mention that while the use of network models has been often overlooked due to the difficulty to monitor and define the interactions in real communities, in the authors' opinion the conception of more advanced tracking systems during the last pandemic leads naturally to this kind of approaches. 

The problem of estimating the state of partially observable dynamic networks has been object of only a few studies in the last few years. One of the most studied problems is the estimation of the source of an information spread in networks using only limited observations. \cite{zhu_information_2013,zhu_robust_nodate} propose a sample path algorithm to estimate the location of a source of information or a disease. \cite{alexandru_diffusion_2019} extend this idea to the case where multiple rumours are spread and the time of the origin of the information is unknown. These works provide interesting idea that can be possibly adapted for the estimation of the evolution of an epidemic over a network.

An alternative approach to the surveillance of epidemics within networks can be found on the use of a sentinel system to estimate the evolution of the epidemic as done by~\cite{braeye_incidence_2019}. A sentinel system involves a limited network of selected reporting sites monitoring the disease in small portions of the population. The obtained data is used to estimate the behaviour of the entire network. \cite{souty_improving_2016} estimate the total number of cases of influenza based on the population density associated to each reporting site. Although this approach uses the density of population to improve the estimation of the state of the epidemic, the total population is still divided into small clusters with homogeneous distribution and interactions.

At the time of the writing of this paper, some early work presenting attempts to define smart testing and quarantine policies have been just published. In particular \cite{berger_seir_nodate} propose a policy based on conditional quarantine and random testing. However, the model based on partial observations assumes that tested negatives are "tagged" and they remain observable after a single test. In another recent paper on the subject by~\cite{kasy_adaptive_2020}, the trade-off between quarantine and testing is  regarded by defining a certain threshold based on the infection probability and related to the cost of testing or quarantining an individual. In this case, the partial information is inferred based on the social group of the individual rather than its interactions within the network. 

The main contribution of this paper is to propose a smart testing strategy to select the individuals to be tested based on the estimated probability of infection of each individual. As a first step we propose a method to make an approximated estimation of the current state of the epidemic which is computationally inexpensive. On the basis of this estimation, the testing policy is defined as a constrained optimization problem. This testing policy is coupled with a selective confinement policy which allows to only confine few individuals of the population based on the outcome of the tests. 
We compare the proposed strategy with the current best practice, namely contact tracing of positive, and a suitable topology-based strategy, where individuals to test are selected according to their number of contacts.
Numerical simulations show the advantage of this approach both in terms of number of infected individuals and in terms of number of individuals put in quarantine at each time. 
In particular, on a population of 10'000 individuals, the total number of infected is 8 times less and the total amount of days spent in quarantine is 5 times less with respect to the current best practice, and the improvement with respect to the topology-based strategy is even more evident.
These results also show that tracing of contacts is crucial to keep under control the epidemic but it can be largely improved by using the algorithms proposed in this paper.
The proposed algorithms can be used in a centralized way (e.g. by a decision maker) but they are also suitable to work in a distributed privacy-aware fashion and to integrate with tracing devices.

The remainder of the paper is organized as follows. In Section~\ref{sec:model_derivation} the proposed model of the epidemic is presented. Sections~\ref{sec:exact_solution} and \ref{sec:approx_estimation} introduce the exact and the approximated estimations of the evolution of the epidemic. Section~\ref{sec:testing} defines the testing strategy and Section~\ref{sec:quarantine}  the quarantine actions. In Section~\ref{sec:numerical_simulations} several simulations demonstrate the performance of the proposed strategies.  Section~\ref{sec:conclusions} provides conclusions and future works.

\section{Model}\label{sec:model_derivation}

Consider a population of $N$ individuals where a disease is spreading.
Each individual can be susceptible, infected, or removed. The spreading of the epidemic is modelled according to the following assumption.
\begin{assumption}
A susceptible individual can be infected by other infected individuals of the population with whom he had a direct contact. Once an individual is infected, the individual will eventually become removed and cannot be infected a second time.
\end{assumption}
The exposure to an infected individual is a necessary but not-sufficient condition for a susceptible individual to become infected. Indeed, the contagion actually takes place if some events (e.g. exchange of body fluids for flu-like illnesses) have occurred and thus it is intrinsically stochastic. Motivated by these considerations, we model the transmission of the disease through random variables. Similarly, also recovery is modelled as a random variable to capture the uncertainty of the recovery process. 

Mathematically, each individual $i$ has at fixed time instants, say every day, $t$ a state $\xi_i(t)\in \{S,I,R\}$ that can take three logical states: 
\begin{itemize}
\item S - susceptible, the individual is healthy and was never infected before, so it is susceptible of being infected; 
\item I - infected, the individual is infected and can infect others;
\item R - removed, the individual has been infected in the past and cannot be infected anymore (because immune or dead).
\end {itemize}
We denote with $u_i(t)\in\{0,1\}$ the binary stochastic input representing the stochastic contagion event at time $t$. The variable takes value $u_i(t)=1$ if the $i$th individual has been infected between time $t$ and time $t+1,$ and $u_i(t)=0$ otherwise.

To characterize $u_i$ we introduce the transmission variables $T_{ji}(t)\in\{0,1\}$ which takes value $T_{ji}(t)=1$ if the infection is transmitted from $j$ to $i$ between time $t$ and time $t+1$ \textit{given that} the individual $j$ was infected and the individual $i$ was susceptible. The same way $r_i(t)\in\{0,1\}$ denotes the binary stochastic variable representing the stochastic recovery event at time $t$. In particular, $r_i(t)=1$ if the $i$th individual becomes removed between time $t$ and time $t+1,$ and $r_i(t)=0$ otherwise. Note that the recovery variable $r_i(t)$ denotes the moment that an individual is not infected anymore, due to immunity or death.

Finally the state of each individual evolves according to the following equation
\begin{equation}
\xi_i(t+1)=
\begin{cases}
S &\text{if } \xi_i(t)=S \text{ and } u_i(t)=0\\
I &\text{if } \xi_i(t)=S \text{ and } u_i(t)=1 \\ 
&\quad \text{ or if } \xi_i(t)\!=\!I \text{ and } r_i(t)=0\\
R &\text{if } \xi_i(t)=I \text{ and } r_i(t)=1 \\ &\quad \text{ or if }  \xi_i(t)=R
\end{cases}
\end{equation}
with
\begin{equation}
u_i(t) = 1 - \prod_{j\, : \, \xi_j(t)=I} (1-T_{ji}(t))
\end{equation}

\begin{figure}[t]
	\centering
	\includegraphics[width=0.7\columnwidth]{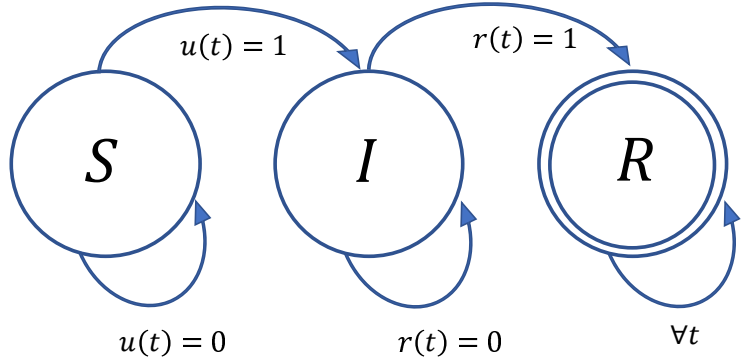}
	\caption{Evolution of the state.\label{fig:automata}}
\end{figure}

The state evolution of each individual is depicted by Fig~\ref{fig:automata}.
From the last equation it is clear that an individual $i$ can be infected by individual $j$ if individual $j$ was infected, i.e. $\xi_j(t)=I$, and if the transmission occurred, i.e. $T_{ji}(t)=1$.
The modelling of variable $T_{ji}(t)$ is summarized by the following assumption.
\begin{assumption}
	The transmission of the disease $T_{ji}(t)$  from an infected individual $j$ to a susceptible individual $i$ is a Bernoulli random variable with mean $w_{ij}(t)$, such as $T_{ij}\sim \mathcal{B}(w_{ij})$. Moreover $T_{ji}(t)$ is independent of $T_{mn}(k) \ \forall m,n,k\neq i,j,t$ and of the initial state $\xi_n(0)\ \forall n $. The mean values are symmetric, i.e. $w_{ij}(t)=w_{ji}(t)$. For any pair $i,j$ of individuals that have no contacts $w_{ij}(t)=0$.
\end{assumption}
The variable $r_i(t)$ is modelled according to the following assumption,
\begin{assumption}
	The recovery $r_i(t)$ is a Bernoulli random variable with mean $\lambda_i$ constant over time. 
	Moreover $r_{i}(t)$ is independent of $r_{j}(k) \ \forall j,k\neq i,t$, of $T_{mn}(k) \ \forall m,n,k$, and of the initial state $\xi_n(0)\ \forall n$.
\end{assumption}

In general the system is partially observable as symptoms only appear in a small percentage of the population.
The appearance of symptoms is modeled by the random variable $e_i(t) \in \{0,1\}$ taking value $e_i(t)=1$ if $i$-th individual is infected and shows symptoms between time $t$ and time $t+1$, and $0$ otherwise. We model it according to the following assumption.
\begin{assumption}
	The appearance of symptoms $e_i(t)$ is a Bernoulli random variable with mean $\theta_i$ constant over time. Moreover $e_{i}(t)$ is independent of $e_{j}(k)$ and $r_{j}(k)\ \forall j,k\neq i,t$, of $T_{mn}(k) \ \forall m,n,k$, and of the initial state $\xi_n(0)\ \forall n$.
\end{assumption}

\subsection{Problem formulation}

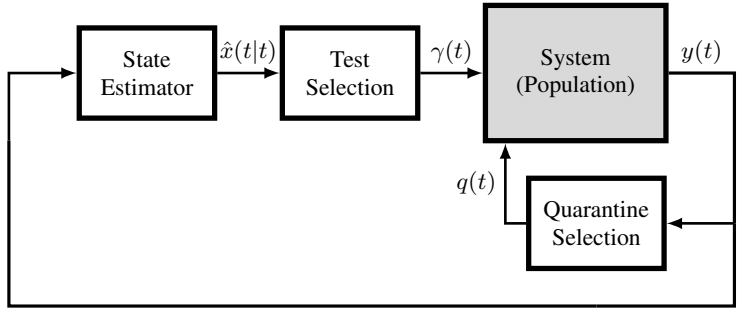
\begin{figure}[t]
	\centering
	\begin{tikzpicture}
	\tikzstyle{every node}=[font=\small]
	
	\pgfmathsetmacro{\w}{1.8}
	\pgfmathsetmacro{\h}{1.2}
	
	\coordinate (est) at (0,0);
	\coordinate (test) at (2.7,0);
	\coordinate (sys) at (5.7,0);
	\coordinate (ctr) at (5.975,-2);
	
	\node[draw, minimum width=\w cm, minimum height=\h cm, line width=2pt, align=center] (EST) at (est) {State \\ Estimator};
	\node[draw, minimum width=\w cm, minimum height=\h cm, line width=2pt, align=center] (TEST) at (test) {Test \\ Selection};
	\node[draw, minimum width=2.4 cm, minimum height=1.8 cm, line width=2pt, align=center, fill=white!85!black] (SYS) at (sys) {System \\ (Population)};
	\node[draw, minimum width=\w cm, minimum height=\h cm, line width=2pt, align=center] (CTR) at (ctr) {Quarantine \\ Selection};
	
	\coordinate (in) at ($(EST.west)-(\w/2,\w/2)$);
	\coordinate (out1) at ($(CTR.east)+(\w/2,0)$);
	\coordinate (out2) at ($(CTR)-(0,1.1)$);
	
	\draw[-latex, line width=1pt] (EST) -- (TEST) node[pos=0.5, above]{$\hat{x}(t|t)$};
	\draw[-latex, line width=1pt] (TEST) -- (SYS) node[pos=0.5, above]{$\gamma(t)$};
	\draw[-latex, line width=1pt] (in) |- (EST);
	\draw[-latex, line width=1pt] (out1) -- (CTR);
	\draw[-, line width=1.2pt] (SYS) -| (out1) node[pos=0.25, above]{$y(t)$};
	\draw[-, line width=1.2pt] (out1) |- (out2);
	\draw[-, line width=1.2pt] (out2) -| (in);
	\draw[-latex, line width=1pt] (CTR.west) -| ($(SYS.south west)+(\w/6,0)$) node[pos=0.75, left]{$q(t)$};
	\end{tikzpicture}
	\caption{Control scheme}
	\label{fig:ctrlscheme}
\end{figure}

In this paper we consider the case in which only a limited amount $N_T$ of tests are available at each time $t$. 
We assume that when a test is performed on the $i$-th individual at time $t$ we obtain the information if $\xi_i(t)=I$ or not. No other information is provided by the test, so it is not possible to distinguish if an individual is susceptible or recovered. We can formally introduce the auxiliary state $x_i(t)$, taking value $x_i(t)=1$ if $\xi_i(t)=I$, and $x_i(t)=0$ otherwise, that represents the binary variable accessed by the test.
For Covid-19 the value of $x_i(t)$ can be retrieved by exploiting several different kind of tests. To date, even if false negative are not completely avoided, PCR tests are widely considered to be very accurate (it is the only recommended method for the identification of Covid-19, see \cite{who2020advice}, and the reference standard for many medical studies, see e.g. \cite{dinnes2020rapid}). Based on this consideration, the paper assumes that available tests are ideal which is in line with an ample part of the literature dealing with testing strategies~\citep{piguillem_optimal_nodate,berger_seir_nodate}.

To model the testing phase, we introduce the selection variables $\gamma_i(t)$ taking value $\gamma_i(t)=1$ if individual $i$ is selected to be tested at time $t$ and $\gamma_i(t)=1$ otherwise. Variable $\gamma$ is thus a controlled variable that can be managed to tackle the disease diffusion. Beside tested individuals, we also consider that additional information is provided by symptomatic individuals. In the context of this work, a symptomatic individual is assumed to be an infected person who spontaneously visits a medical center with clear symptoms and it is diagnosed with the disease.
Let $\mathcal{S}(t) = \{s_1(t),\,s_2(t) \,,\, \dots \,,\, s_{M(t)}(t)\}$ be the set containing the indices of the individuals who are tested at the time instant $t$ and of the individuals who show first symptoms at time $t$. Note that the cardinality $M(t)$ of the set is time-dependent since the number of symptomatic individuals is not constant.
The observed output at time $t$ can be then expressed as
\begin{equation}
y(t) = \left\{x_{s_1(t)}(t),\,x_{s_2(t)}(t),\, \dots \,,\, x_{s_{M(t)}}(t) \right\}
\end{equation}
while the set of the observed outputs up to time instant $t$ is
\begin{equation}
Y^{0:t} = \left\{y(0),\, y(1),\, \dots \,,\, y(t) \right\}
\end{equation}
and it represents the information available at time instant $t$. 
Beside the testing phase, the system evolution is affected by the quarantine mechanism represented by the control variable $q_i(t)$, taking value $q_i(t)=1$ if individual $i$ is in quarantine at time $t$ and $q_i(t)=0$ otherwise. The variable $q(t)$ is  the control variable that governments use to tackle the epidemic.

The goal of this paper is the definition of a policy to select the individuals to be tested that, in conjunction with a selective quarantine policy, is able to reduce the spread of the disease while keeping a limited number of individuals in quarantine.
To do so, we tackle the problem by proposing the closed-loop structure reported in Fig.\ref{fig:ctrlscheme} consisting of three stages: 1) estimation of the states $x_i(t)$ using the information available $Y^{0:t}$ from the feedback of the outputs; 2) selection of the $N_T$ individuals to test by optimizing a reward function $R(\cdot)$; and 3) based on the output $y(t)$, execution of control actions  through selective quarantine. The following sections focus on the derivation of a proper state estimation given the information available $Y^{0:t}$ and the definition of a suitable reward.

\section{Exact estimate: Hidden Markov Model}\label{sec:exact_solution}

The state of the whole population is defined by the vector
\begin{equation}
\xi(t) = \left( \xi_1(t),\, \xi_2(t),\,\dots\,,\, \xi_N(t)\right)\in \{S,I,R\}^N.   
\end{equation}
Under Assumptions 1,2,3, at any time $t$, the next state of the population $\xi(t+1)$ depends only on the current state of the population $\xi(t).$ Accordingly, in line of principle, the stochastic process describing the evolution of the epidemic satisfies the Markov property and can be represented by a Markov chain. 

To model the dynamics of the Markov chain, we have to derive the transition matrix $A \in 3^N\times3^N $ whose entries are
\begin{equation}
A_{\mathfrak{v}\mathfrak{z}} = P(\xi(t+1) = z | \xi(t)=v).
\end{equation}
where $z,v \in \{S,I,R\} ^ N$ represent two possible states of the network, and $\mathfrak{z},\mathfrak{v}$ represent the indices of the transition matrix associated to them. 
Under Assumptions 1,2,3, the next states of two individuals are independent given the previous state of the population. It follows that
\begin{equation}
P(\xi(t+1)=z| \xi(t)=v)=\prod_{i=1}^N P(\xi_i(t)=z_i| \xi(t-1)=v),
\end{equation}
allowing to compute the transition probabilities of the network as a derivation of the transition probabilities between the states of each single individual. 
Since only the transition from susceptible to infected depends on the state of other individuals, the following simplification holds
\begin{equation}
P(\xi_i(t+1)=z_i| \xi(t)=v)= 
\begin{cases}
P(\xi_i(t+1)=z_i| \xi(t)=v) 	&\text{ if } v_i=S\\
P(\xi_i(t+1)=z_i| \xi_i(t)=v_i) &\text{ otherwise}, 
\end{cases}  
\end{equation}
where $z_i,v_i \in \{S,I,R\}$.

The state transition probability of any individual at time $t$ can be computed as
\begin{equation} 
\begin{cases}
P(\xi_{i}(t+1) =  I| \xi(t)=v,\, v_i=S) = 1 - \prod_{j \,:\, v_j=I} (1-w_{ji}(t)) \\
P(\xi_{i}(t+1) = R | \xi_{i}(t) = I) = \lambda_i,
\end{cases}
\end{equation} 
while the probability of remaining in a given state is
\begin{equation} 
\begin{cases}
P(\xi_{i}(t+1) = S |  \xi(t)=v,\, v_i=S) = 1 - P(\xi_{i}(t+1) = I |  \xi(t)=v,\, v_i=S), \\
P(\xi_{i}(t+1) = I | \xi_{i}(t) = I) = 1 - P(\xi_{i}(t+1) = R | \xi_{i}(t) = I), \\
P(\xi_{i}(t+1) = R | \xi_{i}(t) = R) = 1.
\end{cases}
\end{equation} 
All other transitions have probability 0.

A major difficulty in our setting is that the evolution of the system can be only observed by symptomatic individuals and selective tests on the population. Since $M(t) <N$, the Markov model is hidden and can be only partially observed through the output. The complete characterization of the state given the available information is provided by the joint distribution $p(\xi(t), Y^{0:t})$. For a given $Y^{0:t}$, the joint distribution can be represented by a vector of dimension $3^N$ with entries $p(z, Y^{0:t})= \hat{p}_t(z)$ $\forall z \in \{S,I,R\}^N$. In the case of Hidden Markov Models, this joint distribution can be easily computed by means of the \textit{forward algorithm}~(see \cite{blunsom2004hidden,l_r_rabiner_tutorial_1989}), providing the following expression
\begin{align}
\hat{p}_t(z) 
&= p(\xi(t)=z, \, Y^{0:t-1},\, y(t))  \nonumber  \\
&= P(y(t)|\xi(t)=z, \, Y^{0:t-1}) p(\xi(t)=z,\, Y^{0:t-1})  \nonumber  \\
&= P(y(t)|\xi(t)=z)  \sum\nolimits_{v} P(\xi(t) = z | \xi(t\!-\!1)=v) p(\xi(t\!-\!1)=v, \, Y^{0:t-1}) \nonumber  \\
&= P(y(t)|\xi(t)=z) \sum\nolimits_{v} A_{\mathfrak{v}\mathfrak{z}}  \hat{p}_{t-1}(v).
\end{align}
The computation of $P(y(t)|\xi(t)=z)$ is then easy: $P(y(t)|\xi(t)=z)=1$ if the state $\xi(t)=z$ gives the output $y(t)$, and 0 otherwise, namely if $y(t)$ is not a possible output for the state $z$.

From the joint distribution, the conditional distribution is
\begin{align}
p(\xi(t)=z\,|\, Y^{0:t})
=\frac{p(\xi(t)=z,\, Y^{0:t})}{p(Y^{0:t})}
=\frac{p(\xi(t)=z,\, Y^{0:t})}{\sum_{z}p(\xi(t)=z,\, Y^{0:t})}
\end{align}
where we used the Bayes rule and the law of total probability. Recall now that the optimal estimate of a random variable corresponds to the expected value given the observations, see \cite{anderson2012optimal}. Then the optimal estimate of the $i$-individual $\hat{x}_i(t|t)$ is 
\begin{align}
\hat{x}_i(t|t)
=\mathbb{E} \left[x_i(t)|Y^{0:t}\right]=P(x_i(t)\!=\!1|Y^{0:t})
=\sum_{z:z_i=I} p(\xi(t)\!=\!z|Y^{0:t}).
\end{align}
where we used the definition of expected value for a binary variable and the law of total probability.
This procedure allows to obtain the probability of each individual to be infected at time $t$ given the complete vector of observations $Y^{0:t}$. However, in spite of allowing to compute the exact probability, this approach requires the computation of all the transition probabilities of the matrix $A$ and the use of a vector variable of size $3^N$, which is not computationally feasible even for small populations.

\section{An approximated state estimate}\label{sec:approx_estimation}

Due to the prohibitive burden of an exact probability computation, in this paper we propose an approximated low-computational algorithm to estimate such probability.  The proposed approximated estimation is based on the idea of temporal and spatial truncation of the updates and is also suitable for  decentralized implementations
More precisely, we propose to propagate the information from testing only to individuals that are the most correlated to the tested individuals, namely the ones that have direct contact with tested people, while for the remaining part of the population the update is performed based on previous estimates and the topology of the network representing the population.  In the same way, only a limited amount of past state estimates are assumed to be affected by the new information. This approximation allows to retrieve the information regarding the  individuals that are most affected by the result of each test while keeping a limited computational time.

\subsection{Approximated State Estimation Definitions}

We define the estimate of the state of the individual $i$  as
\begin{equation}
\hat{x}_i(t|t)=\mathbb{E}[x_i(t)\,|\,\mathcal{I}_i^{0:t}]=P(x_i(t)\!=\!1\,|\,\mathcal{I}_i^{0:t}) \end{equation}
where the local information set for the $i$-th individual $\mathcal{I}_i^{0:t}$ is defined as
\begin{multline}
\mathcal{I}_i^{0:t}= \mathcal{I}_i^{0:t-1}\cup \left\{\hat{x}_k(t-1|t-1), \ k\,:\, w_{ik}(t)\neq0 \right\} 
\\ \cup \left\{\hat{x}_k(\tau|t), \ \tau\leq t-1,\, k\in \mathcal{S}(t)\right\} \cup y(t).
\end{multline}
Local information set $\mathcal{I}_i^{0:t}$ consists of the local state estimates of direct contacts, updated at the time instant of the interaction, and the state estimates of tested individuals, updated just after the test.
In order to keep the computations associated to the $i$-th individual limited, we define a local approximated estimation which can be retrieved based only a partial knowledge of the network.
More precisely, the state of each individual is estimated under the assumption that only its contacts are known, and no information on the connections between any other individuals is assumed to be available.
In this sense, we will focus only on individuals with whom
the  $i$-th individual was in contact: in the case of untested individuals we will use only the previous local estimates $\hat{x}_k(t-1|t-1)$ while in the case of tested individuals we will use the current state ${x}_k(t)\in y(t)$ and the updated estimation $\hat{x}_k(\tau|t), \ \tau\leq t-1$ of past states.

Denote by $X_i^{0:t}$ the (random) vector collecting all the random variables $x_i(\tau) \ \tau\leq t$, namely $X_i^{0:t}=\left(x_i(0), \,\dots\,,\, x_i(t)\right)'$. With a little misuse of notation $X_i^{0:t}=0$ denotes the case where all the past states $x_i(\tau), \ \tau\leq t,$ are equal to $0$.
We assume that the random variables $T_{ji}(\tau)x_j(\tau),\ j\neq i,$ are conditional independent given $X_i^{0:\tau}=0$. Similar assumptions are made by \cite{valdano2015analytical} and \cite{boguna2013nature}. The rationale is that, if the individual $i$ has always been healthy, the coupling between two of his neighbours $m$ and $n$ is negligible, as in the case when individual $i$ is the only connection between $m$ and $n$ or, even if $w_{mn}(\tau)\neq0$, contacts of $n$ are enough different from contacts of $m$. It follows that
\begin{equation}\label{eq:ass:independece}
\mathbb{E}\left[u_i(\tau) \, |\, X_i^{0:\tau}=0\right]
=1-\prod_{j=1}^N 1 - w_{ij}(\tau) P(x_j(\tau)\!=\!1 | X_i^{0:\tau}\!=\!0).
\end{equation}

To further simplify the estimation algorithm we will simplify the stochastic recovery with a deterministic one based on the average recovery time $D$. Then we have
\begin{align}
\hat{x}_i(t|t)\label{eq:ass:det_rec}
&=P(x_i(t)\!=\!1\,|\,\mathcal{I}_i^{0:t}) \nonumber \\
&= P(\text{at least a contagion in }(t\!-\!D,t) \cap X_i^{0:t-D}=0\,|\,\mathcal{I}_i^{0:t})  \nonumber \\
&= P(X_i^{0:t-D}=0\,|\,\mathcal{I}_i^{0:t})  - P(X_i^{0:t}=0\,|\,\mathcal{I}_i^{0:t}) 
\end{align}
We compute $P(X_i^{0:t}\!=\!0\,|\,\mathcal{I}_i^{0:t})$ and $P(X_i^{0:t-D}\!=\!0\,|\,\mathcal{I}_i^{0:t})$ as
\begin{align}
P(&X_i^{0:\tau}=0\,|\,\mathcal{I}_i^{0:t}) \nonumber \\
&=P(u_i(\tau\!-\!1)=0 \, \cap \, X_i^{0:\tau\!-\!1}=0\,|\,\mathcal{I}_i^{0:t}) \nonumber \\
&=P(u_i(\tau\!-\!1)=0 \,|\, X_i^{0:\tau\!-\!1\!}=0,\,\mathcal{I}_i^{0:t})P(X_i^{0:\tau\!-\!1}=0\,|\,\mathcal{I}_i^{0:t}) \nonumber \\
&=\left[1-\mathbb{E}\left[u_i(\tau\!-\!1) \,\Big|\, X_i^{0:\tau\!-\!1\!}=0,\,\mathcal{I}_i^{0:t}\right]\right] P(X_i^{0:\tau\!-\!1}=0\,|\,\mathcal{I}_i^{0:t}) \nonumber \\
&=\left[\prod_{j=1}^N \!1 - w_{ij}(\tau\!-\!1) P(x_j(\tau\!-\!1)\!=\!1 | X_i^{0:\tau\!-\!1}\!=\!0,\mathcal{I}_i^{0:t})\right]\!P(X_i^{0:\tau\!-\!1}\!=\!0|\mathcal{I}_i^{0:t})
\end{align}
where the last equality holds since $T_{ji}(\tau)x_j(\tau)\ j\neq i$ are conditional independent given $X_i^{0:\tau\!-\!1}=0$.
To obtain the numerical value of $P(X_i^{0:\tau}=0\,|\,\mathcal{I}_i^{0:t})$ would require to compute $P(x_j(\tau\!-\!1)=1 \, | \, X_i^{0:\tau\!-\!1}=0,\,\mathcal{I}_i^{0:t})$ that in turn would require $P(x_k(\tau-2)=1\,| X_j^{0:\tau\!-\!1}=0,\, X_i^{0:\tau\!-\!1}=0,\,\mathcal{I}_i^{0:t})$ and so on. 
Since this propagation is very computationally expensive we  make the approximation that
\begin{align}\label{eq:ass:selfcont1}
P(x_j(\tau\!-\!1)\!=\!1 | X_i^{0:\tau\!-\!1}\!=\!0,\mathcal{I}_i^{0:t}) = P(x_j(\tau\!-\!1)=1 \, | \,\mathcal{I}_i^{0:t}).
\end{align}
The underlying assumption is that the state of an individual and those of its neighbors are independent. The accuracy of this assumption has been explored by \cite{gleeson2012accuracy} where it has been shown that the dynamics are well approximated if the degrees of closest neighbours are high. 
In the same way the assumption holds when the underlying network of contacts is time-varying, the results can be less accurate if pairs of individuals have frequent interactions, and many contacts in common. Since this happens in real life (think of relatives and colleagues), we introduce a correction factor $\sigma_{ij}(w_{ij}(\tau),w_{ij}(\tau\!-\!1), \dots, w_{ij}(0)))\in [0,1]$, for simplicity denoted by $\sigma_{ij}(\tau)$, that accounts for the coupling of individuals $i$ and $j$ due to the interactions before $\tau$
\begin{align}\label{eq:ass:selfcont2}
P(x_j(\tau\!-\!1)\!=\!1 | X_i^{0:\tau\!-\!1}\!=\!0,\mathcal{I}_i^{0:t}) = \sigma_{ij}(\tau\!-\!1)P(x_j(\tau\!-\!1)=1 \, | \,\mathcal{I}_i^{0:t}).
\end{align}
In line of principle, $\sigma_{ij}(\tau)$ is as smaller as many interactions have occurred between $i$ and $j$ in the past. In fact, the probability that $j$ is infected given that $i$ has been healthy (namely the left hand side of eq. \eqref{eq:ass:selfcont1}) is lower than the probability that $j$ is infected without any knowledge on the past states of $i$ (namely the right hand side of eq. \eqref{eq:ass:selfcont1}). An efficient way to compute $\sigma_{ij}(\tau)$ is defined in Sec. \ref{sec:numerical_simulations}.
We can conveniently incorporate the correction factor  $\sigma_{ij}(\tau)$ in the term $w_{ij}(\tau)$ as $\bar{w}_{ij}(\tau)=c_{ij}(\tau)w_{ij}(\tau)$.
We finally obtain the following update rule
\begin{equation}\label{eq:ass:selfcont}
P(X_i^{0:\tau}=0\,|\,\mathcal{I}_i^{0:t}) =  \left[\!\prod_{j=1}^N \!1 \!-\! \bar{w}_{ij}(\tau\!-\!1) P(x_j(\tau\!-\!1)\!=\!1 | \mathcal{I}_i^{0:t})\!\right]\!P(X_i^{0:\tau\!-\!1}\!=\!0|\mathcal{I}_i^{0:t}).
\end{equation}

To keep a limited number of computations, we also make the following approximation
\begin{equation}\label{eq:ass:spacetrunc}
P(x_j(\tau)=1\,|\, \mathcal{I}_i^{0:t}) =
\begin{cases}
P(x_j(\tau)=1\,|\, \mathcal{I}_j^{0:t}) &{\text{ if } j\in \mathcal{S}(t)} \\
P(x_j(\tau)=1\,|\, \mathcal{I}_i^{0:t-1})  &\text{ otherwise.}
\end{cases}
\end{equation}
with initialization $P(x_j(\tau)=1\,|\, \mathcal{I}_i^{0:\tau})=\hat{x}_j(\tau|\tau)$.
Roughly speaking, if individual $j$ has direct contact with a tested individual $k$ and individual $i$ has direct contact with $j$ but not with $k$, the state estimates of $j$ will be corrected based on the outcome while the state estimates of $i$ will use the old estimation of $j$, as derived without the knowledge of the outcome. 
This means that we use the information regarding the outcome from the tests to only update the direct contacts of tested individuals. 

\subsection{State estimation update}
Since the update of each individual uses only knowledge from local connections, new information can be used differently for tested individuals, individuals with a direct contact with them, and the remaining of the population.

\subsubsection*{Tested individuals}
Let $\gamma$ denote the outcome of the test to the individual $i$. Then we have
\begin{align}
&P(X_i^{0:\tau}=0\,|\, \mathcal{I}_i^{0:t}) = P(X_i^{0:\tau}=0|\, x_i(t)=\gamma,\, \mathcal{I}_i^{0:t-1})\\
&P(x_i(\tau)=1\,|\, \mathcal{I}_i^{0:t}) = P(x_i(\tau)=1\,|\, x_i(t)=\gamma, \, \mathcal{I}_i^{0:t-1}).
\end{align}
If $x_i(t)=0$, no contagion happened in $(t,t-D)$, namely
\begin{equation}
P(X_i^{0:\tau}=0\,|\, x_i(t)=0, \mathcal{I}_i^{0:t-1}) = P(X_i^{0:t-D}=0\,|\,\mathcal{I}_i^{0:t-1}) 
\end{equation}
for $\tau=t-D+1,\dots\,,\, t$, and 
\begin{equation}
P(X_i^{0:\tau}=0\,|\, x_i(t)=0, \mathcal{I}_i^{0:t-1}) = P(X_i^{0:\tau}=0\,|\,\mathcal{I}_i^{0:t-1}) 
\end{equation}
for $\tau\leq t-D$ as no additional information on past states is given by a negative outcome. As $x_i(t)=0$, $x_i(\tau)$ may be equal to 1 only if a contagion occurs in the interval $(\tau-D, t-D)$, therefore the infection probability is updated as
\begin{align}\label{eq:est:testedneg}
\hat{x}_i(\tau|t)
&=P(x_i(\tau)=1\,|\, x_j(t)=0,\, \mathcal{I}_i^{0:t-1}) \nonumber\\
&= \frac{P(x_i(\tau)=1 \cap x_i(t)=0 \,|\, \mathcal{I}_i^{0:t-1})}{P(x_i(t)=0\,|\, \mathcal{I}_i^{0:t-1}) }\nonumber\\
&= \frac{P(X_i^{\tau-D}=0\,|\, \mathcal{I}_i^{0:t-1}) - P(X_i^{t-D}=0\,|\, \mathcal{I}_i^{0:t-1})}{P(x_i(t)=0\,|\, \mathcal{I}_i^{0:t-1})}
\end{align}
for $\tau=t-D+1,\dots\,,\, t$ and 
\begin{equation}
\hat{x}_i(\tau|t) = P(x_i(\tau)\!=\!1\,|\, x_j(t)\!=\!0,\, \mathcal{I}_i^{0:t\!-\!1}) = P(x_i(\tau)\!=\!1\,|\, \mathcal{I}_i^{0:t\!-\!1}) = \hat{x}_i(\tau|t\!-\!1)
\end{equation}
for $\tau\leq t-D$.

For the case of a positive result, $x_i(t)=1$, we have
\begin{equation}
P(X_i^{0:\tau}\!=\!0 \,|\, x_i(t)\!=\!1, \mathcal{I}_i^{0:t\!-\!1}) = 1
\end{equation}
for $\tau\leq t-D$, while for $\tau=t-D+1,\dots\,,\, t$ it holds that
\begin{align} \label{eq:est:testedpos}
P(&X_i^{0:\tau}=0\,|\, x_i(t)=1, \mathcal{I}_i^{0:t-1}) \nonumber \\
&=\!P(X_i^{0:t\!-\!D}\!=\!0 \cap \text{no contagion in } (t\!-\!D,\tau) \,|\, x_i(t)\!=\!1, \mathcal{I}_i^{0:t\!-\!1}) \nonumber \\
&=\!1\!-\!P(\text{at least a contagion in}\, (t\!-\!D,\tau) | X_i^{0:t\!-\!D}\!=\!0, x_i(t)\!=\!1, \mathcal{I}_i^{0:t\!-\!1}) \nonumber \\
&=\!1\!-\!\frac{P(\text{at least a contagion in}\, (t\!-\!D,\tau) \cap x_i(t)\!=\!1 | X_i^{0:t\!-\!D}\!=\!0, \mathcal{I}_i^{0:t\!-\!1})}{P(x_i(t)\!=\!1 \,|\,X_i^{0:t-D}\!=\!0, \mathcal{I}_i^{0:t-1})} \nonumber \\
&=1-\frac{P(\text{at least a contagion in } (t-D,\tau) \,|\,X_i^{0:t-D}\!=\!0,\, \mathcal{I}_i^{0:t-1})}{P(x_i(t)\!=\!1 \,|\,X_i^{0:t-D}\!=\!0,\, \mathcal{I}_i^{0:t-1})} \nonumber \\
&=1-\frac{P(X_i^{0:t-D}\!=\!0\,|\, \mathcal{I}_i^{0:t-1}) - P(X_i^{0:\tau}\!=\!0 \,|\, \mathcal{I}_i^{0:t-1})}{P(X_i^{0:t-D}\!=\!0\,|\, \mathcal{I}_i^{0:t-1}) - P(X_i^{0:t}\!=\!0 \,|\, \mathcal{I}_i^{0:t-1})}
\end{align}
If $x_i(t)=1$,  $x_i(\tau)$ is equal to 1 only if the contagion occurred in the interval $(t-D, \tau)$. Knowing that $P(X_i^{t-D}=0\,|\, \mathcal{I}_i^{0:t})=1$, we can compute the infection probability for these individuals as
\begin{equation}
\hat{x}_i(\tau|t)=P(x_i(\tau)=1\,|\, x_i(t)=1, \mathcal{I}_i^{0:t-1}) = 1 - P(X_i^{0:\tau}=0\,|\, \mathcal{I}_i^{0:t})
\end{equation}
for $\tau=t-D+1,\dots\,,\, t$ and 
\begin{equation}
\hat{x}_i(\tau|t)=P(x_i(\tau)=1\,|\, x_i(t)=1, \mathcal{I}_i^{0:t-1}) = 0 \end{equation}
for $\tau\leq t-D$.

\subsubsection*{Neighbours of tested individuals}
Formally the neighbours of a tested individual are defined by the set $Q(t)=\{i\,|\, \exists w_{ik}(\tau)\neq0,\ k\in \mathcal{S}(t),\ \tau<t \}$ which represents the set of individuals that has been in contact at least once with at least a tested individual. According to the definition of local information set, the update of the estimation exploits also the updated estimate of the past states of tested individual.

The probability relative to the initial time instant is not changed
\begin{equation}
P(X_i^{0}=0\,|\, \mathcal{I}_i^{0:t}) = P(X_i^{0}=0).
\end{equation}
By using the information from the contacts that have been tested at time instant $t$ we can update the probabilities starting from \eqref{eq:ass:selfcont} as
\begin{align} \label{eq:est:neighbours}
P(X_i^{0:\tau}=0|\mathcal{I}_i^{0:t}) 
&= 
\begin{multlined}[t]
P(X_i^{0:\tau-1}=0|\mathcal{I}_i^{0:t})\cdot
\prod_{j\in \mathcal{S}(t)} 1 \!-\! \bar{w}_{ij}(\tau\!-\!1) P(x_j(\tau\!-\!1)=1|\mathcal{I}_i^{0:t})\cdot\\
\prod_{k\notin \mathcal{S}(t)} 1 \!-\! \bar{w}_{ik}(\tau\!-\!1) P(x_k(\tau\!-\!1)=1|\mathcal{I}_i^{0:t})
\end{multlined} \nonumber \\
&=
\begin{multlined}[t]
P(X_i^{0:\tau-1}=0|\mathcal{I}_i^{0:t})\cdot
\prod_{j\in \mathcal{S}(t)} 1 \!-\! \bar{w}_{ij}(\tau\!-\!1) P(x_j(\tau\!-\!1)=1|\mathcal{I}_j^{0:t})\cdot\\ \quad
\prod_{k\notin \mathcal{S}(t)} 1 \!-\! \bar{w}_{ik}(\tau\!-\!1) P(x_k(\tau\!-\!1)=1|\mathcal{I}_i^{0:t-1}) \hspace{-10pt}
\end{multlined} \nonumber \\
&= P(X_i^{0:\tau}=0\,|\,\mathcal{I}_i^{0:t-1})c_{i,1}(\tau,t)c_{i,2}(\tau,t) 
\end{align}
using \eqref{eq:ass:spacetrunc}, where 
\begin{align}\label{eq:est:neighbours_corr}
c_{i,1}(\tau,t)&=\frac{P(X_i^{0:\tau\!-\!1}=0\,|\,\mathcal{I}_i^{0:t})}{P(X_i^{0:\tau\!-\!1}=0\,|\,\mathcal{I}_i^{0:t-1})} \\
c_{i,2}(\tau,t)&=\frac{\prod_{j\in \mathcal{S}(t)} 1 - \bar{w}_{ij}(\tau-1) P(x_j(\tau\!-\!1)=1 \, | \,\mathcal{I}_j^{0:t})}{\prod_{j\in \mathcal{S}(t)} 1 - \bar{w}_{ij}(\tau-1) P(x_j(\tau\!-\!1)=1 \, | \,\mathcal{I}_i^{0:t-1})}
\end{align}
Note that the previous update takes advantage from the knowledge of the update estimate of the past state of tested individuals.
The last equality holds only if individual $i$ has not been tested before, otherwise $P(X_i^{0:\tau\!-\!1}=0\,|\,\mathcal{I}_i^{0:t-1})$ would be different according to the update relative to a tested individual.
In that case, the correction procedure starts from the instant where the individual was tested.  
The correction procedure works if more than one neighbour have been tested even in different time instants.
Note that $c_{2,i}(\tau,t)>1$ if $x_j(t)=0$ and $c_{2,i}(\tau,t)<1$ if $x_j(t)=1$. Finally, the infection probability at time $t$ is computed as \begin{align}
\hat{x}_i(t|t) = P(x_i(t)\!=\!1\,|\,\mathcal{I}_i^{0:t}) &= P(X_i^{0:t-D}=0\,|\,\mathcal{I}_i^{0:t})  - P(X_i^{0:t}=0\,|\,\mathcal{I}_i^{0:t}). 
\end{align}

\subsubsection*{Open-loop state estimation}
For each individual not having direct contact with any tested individual, the open-loop estimate is computed as $P(X_i^{0:t}=0\,|\,\mathcal{I}_i^{0:t-1})$ based on the previous estimates $P(x_j(t-1)\!=\!1\,|\,\mathcal{I}_i^{0:t-1})$ provided by its contacts as
\begin{align}\label{eq:est:openloop}
P(X_i^{0:t}=0\,|\,\mathcal{I}_i^{0:t}) 
&= \left[\prod_{j=1}^{N} 1 \!-\! \bar{w}_{ij}(t\!-\!1) P(x_j(t\!-\!1)=1 \,| \,\mathcal{I}_i^{0:t-1})\right]P(X_i^{0:t\!-\!1}=0\,|\,\mathcal{I}_i^{0:t}) \nonumber \\
&= \left[\prod_{j=1}^{N} 1 \!-\! \bar{w}_{ij}(t-1) \hat{x}_j(t\!-\!1|t\!-\!1) \right]P(X_i^{0:t\!-\!1}=0\,|\,\mathcal{I}_i^{0:t}).
\end{align}
using \eqref{eq:ass:selfcont} and \eqref{eq:ass:spacetrunc}. Other required values are updated according to 
\begin{align}
&P(X_i^{0:\tau}\!=\!0\,|\, \mathcal{I}_i^{0:t}) \!=\! P(X_i^{0:\tau}\!=\!0\,|\,\mathcal{I}_i^{0:t-1})  \ \qquad \tau\!\leq t-1\\
&P(x_i(\tau)\!=\!1\,|\, \mathcal{I}_i^{0:t}) \!=\! P(x_i(\tau)\!=\!1\,|\, \mathcal{I}_i^{0:t-1}) \qquad \tau\leq t\!-\!1 \\
&\hat{x}_i(t|t)=P(x_i(t)\!=\!1\,|\,\mathcal{I}_i^{0:t}) = P(X_i^{0:t-D}=0\,|\,\mathcal{I}_i^{0:t})  - P(X_i^{0:t}=0\,|\,\mathcal{I}_i^{0:t}).
\end{align}

\subsection{Overall estimation scheme}
\begin{figure}[h]
	\centering
	\includegraphics[width=0.7\columnwidth]{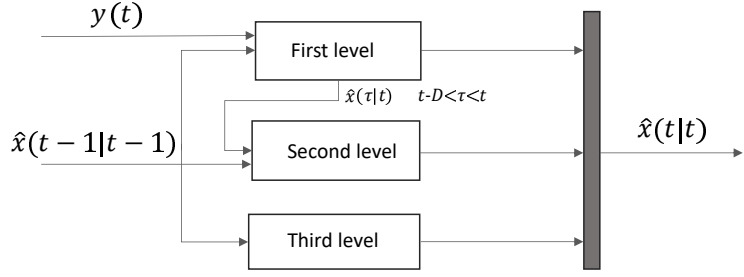}
	\caption{\label{fig:scheme} 3 level update of the state estimate.}
\end{figure}

The state estimation scheme proposed above performs a hierarchical update of the infection probability. This update is structured around individuals that are tested at time $t$, the neighbors of the tested individuals and the remaining of the population. At each time instant, the estimation is thus divided into $3$ levels of update based on the derivations obtained in the previous subsection:
\begin{itemize}
	\item \textit{First level}: Tested individuals, using the output $y(t)$ from the performed tests
	\item \textit{Second level}: Neighbors of tested individuals, including the update estimate from the first level and the previous estimates $\hat{x}(\tau|t), \ \tau<t$
	\item \textit{Third level}: Rest of the population (open loop), using only the previous estimates $\hat{x}(t\!-\!1|t\!-\!1)$.
\end{itemize}
This scheme is depicted in Fig.~\ref{fig:scheme}.

In line of principle, buffers of increasing length are needed to store past probabilities.
In the spirit of a temporal truncation of the updates, since the current test outcomes bring little information on the oldest states except for positive tested individuals, we assume that for untested individuals past probabilities older than $D_w$ are not affected by the new outcomes, i.e.
\begin{equation}
P(X_i^{t:t-D_w}=0\,|\, \mathcal{I}_i^{0:t}) = P(X_i^{t:t-D_w}=0\,|\, \mathcal{I}_i^{0:t-1}) .
\end{equation}
Under this approximation, in terms of information storage, the local update of the current state estimate requires the storage of the following two buffers of information for each individual, namely the Susceptibility buffer
\begin{equation}
B_{sus}^i(t) = \{P(X_i^{0:t-D}=0\,|\, \mathcal{I}_i^{0:t}), \dots, P(X_i^{0:t}=0\,|\, \mathcal{I}_i^{0:t})\}
\end{equation}
and, the Infection probability buffer 
\begin{equation}
B_{inf}^i(t) = \{P(x_i(t\!-\!D\!-\!D_w)=1\,|\, \mathcal{I}_i^{0:t}), \dots, P(x_i(t)=1\,|\, \mathcal{I}_i^{0:t})\}.  
\end{equation}

It is worth to note that the complexity of the proposed estimator is much lower than the optimal estimation devised in the previous section. At each update, the open-loop state estimation requires for each individual the product of a sequence of at most $N$ real factors, see \eqref{eq:est:openloop}. Since only effective contacts (namely those with $w_{ij}\neq0$) affect the product, the number of required multiplications drastically falls and scales with the node degree. The second level update requires, for each contact of a tested individual, the correction of the last $D_w$ values and involves simple multiplications of scalar quantities, see  \eqref{eq:est:neighbours}.
Similarly, at the first level, for each tested individual, the update of the last estimates requires only elementary operations using already available quantities (see \eqref{eq:est:testedneg} and \eqref{eq:est:testedpos}).

\begin{figure}[t]
	\begin{minipage}{\columnwidth}
		\centering
		\includegraphics[width=0.7\columnwidth]{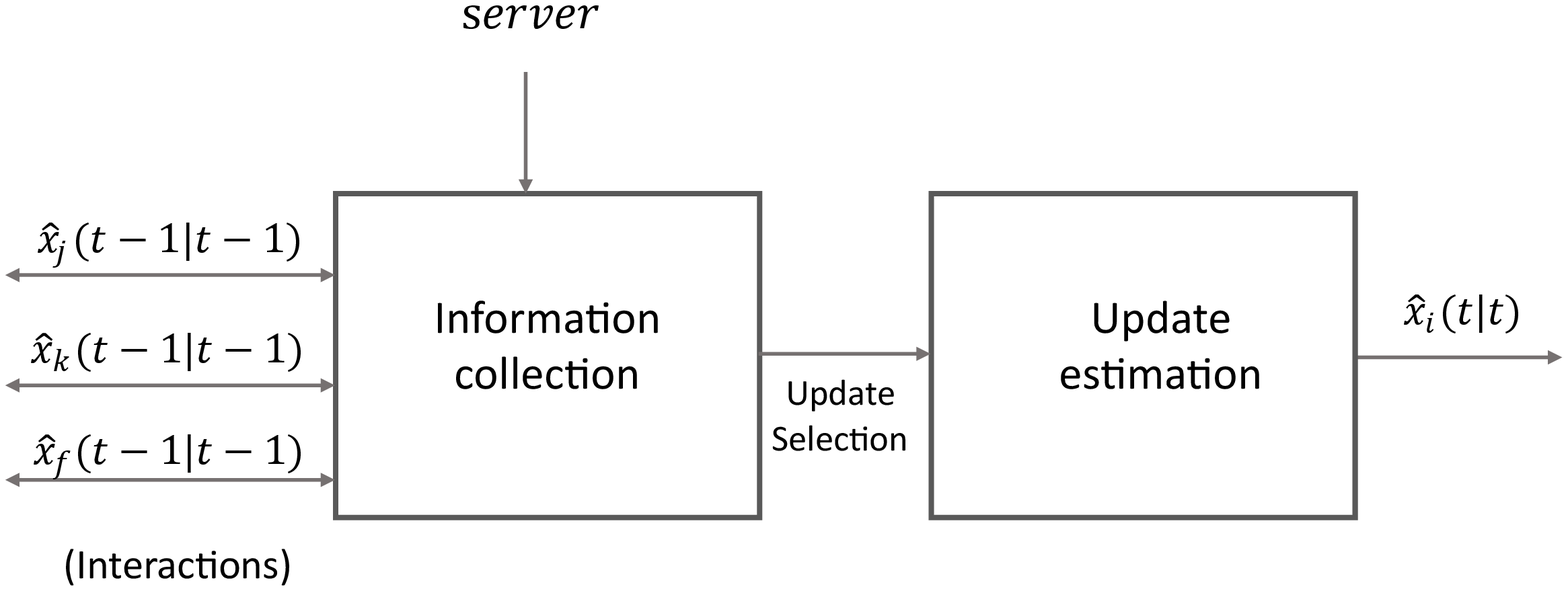}
		\caption{\label{fig:information_update} Information collection and update of each individual.}
	\end{minipage}\vspace*{30pt}
	\begin{minipage}{\columnwidth}
		\centering
		\begin{tikzpicture}
		\coordinate (p0) at (-2.5,-0.8);
		\coordinate (p1) at (0,-0.8);
		\coordinate (p2) at (3.5,0);
		\coordinate (p3) at (6,-1);
		\coordinate (p4) at (3.5,0);
		\pgfmathsetmacro{\l}{0.75};
		
		\node[circle, inner sep=1pt, draw, align=center] (P0) at (p0){\small smartphone \\ \footnotesize (individual k)};
		\node[circle, inner sep=1pt, draw, align=center] (P1) at (p1){\small smartphone \\ \footnotesize (individual i)};
		\node[circle, inner sep=1pt, draw, align=center] (P2) at (p2){\small smartphone \\ \footnotesize (individual j)};
		\node[circle, inner sep=1pt, draw, align=center] (P3) at (p3){\small smartphone \\ \footnotesize (individual l)};
		
		\node[draw, minimum width=1.65cm, minimum height=1.5cm, line width=2pt](server) at (1.5,2.5){\small server};
		
		\draw[latex-latex] (P0) -- (P1);
		\draw[latex-latex] (P1) -- (P2);
		\draw[latex-latex, dotted] (P0) -- (server);
		\draw[latex-latex, dotted] (P1) -- (server);
		\draw[latex-latex, dotted] (P2) -- (server);
		\draw[latex-latex] (P2) -- (P3);
		\draw[latex-] (P1.west) -- ($(P1.west)+(-\l,0)$);
		\draw[latex-] (P1.south west) -- ($(P1.south west)+(-1.1*\l,-0.7*\l)$);
		\draw[latex-] (P1.south) -- ($(P1.south)+(-0.25*\l,-1.25*\l)$);
		\draw[latex-] (P1.south east) -- ($(P1.south east)+(0.5*\l,-1.25*\l)$);
		\draw[latex-] (P2.east) -- ($(P2.east)+(1.25*\l,0.5*\l)$);
		\draw[latex-] (P2.south east) -- ($(P2.south east)+(0.5*\l,-1.25*\l)$);
		\draw[latex-] (P2.south) -- ($(P2.south)+(0,-1.25*\l)$);
		\draw[latex-] (P2.south west) -- ($(P2.south west)+(-\l,-\l)$);
		
		\end{tikzpicture}
		\caption{Required communications. Full lines=local data exchange, dotted lines=remote communication.}
		\label{fig:communications}
	\end{minipage}
\end{figure}

An interesting feature of the proposed approach is that it is not only computationally efficient to be used in a centralized way for a given community, but that it can be also implemented in a decentralized manner. This is the case where each individual is equipped with a smart device (e.g. a smartphone) provided with small computational capabilities and able to communicate with other devices and with a central testing unit, see Fig. \ref{fig:communications}.

Contact tracing mechanisms have already been applied by many countries during Covid-19 epidemic and software are already available in the market. With respect to them, our algorithm can be implemented based on the same hardware and with a larger amount of transmitted data. In particular, when individuals get in contact during the day, their previous estimate $\hat{x}_i(t-1|t-1)$ has to be exchanged.
The outcomes of the tests $y(t)$ are provided to tested individuals who compute and communicate the updated estimates of the previous states $\hat{x}_j(\tau|t), \, j \in \mathcal{S}(t)$ to the server. Then remote communications of those updated estimates are performed once per day from the main server to the population.
An explanatory representation is given in Fig. \ref{fig:information_update}. 

Note that no information on interactions is communicated neither to the central unit or other individuals, so that privacy is preserved and vulnerability of a central data collector is avoided.
Then, each individual transmits the updated estimates $\hat{x}_i(t|t)$ to the server which decides who to test the next day and convoke them.

\section{Testing policy}\label{sec:testing}

Similarly to the literature on sensor selection, it is possible to formulate the test selection problem as a constrained optimization problem based on the state estimate. Formally, we introduce the binary control variable $\gamma_i(t)$ taking value $\gamma_i(t)=1$ if $i$-th individual is selected at time instant $t$ to be tested at the next time instant $t+1$, and $\gamma_i(t)=0$ otherwise, while we denote $\gamma(t)=(\gamma_i(t), \,\dots\,,\, \gamma_N(t))'$. Then the test selection problem can be formulated as
\begin{align}
\gamma_i(t) = &\arg\max_{\gamma}\ R(\hat{x}(t|t),\,\gamma)\\
&\text{s.t. } \sum_{i=1}^{N}\gamma_i \leq N_T
\end{align}
where $R(\cdot)$ is a suitable reward function. 

Several possibilities exist for the choice of the cost function. 
Differently from most of the works on sensor selection for remote estimation, we avoid to adopt the error covariance matrix because it is computationally infeasible for large $N$. More suitable cost functions can be computed based on different metrics of the current state of the population or the topology and characteristics of the network. Namely, intuitive choices would focus on the expected number of detected people, the expected number of infections at the next time instant $t+1$ or targeting individuals with high number of contacts (i.e. first-line health workers). In this context, different cost functions may provide different results based on the time of application of their actions, the number of available tests or the quarantine actions applied.

Based on good preliminary results, in this paper we propose to maximize the expected value of the number of detected positive individuals, that is
\begin{align}\label{eq:cost}
R(\hat{x}(t|t),\,\gamma)
&=\mathbb{E}\left[\sum_{i=1}^{N}\gamma_ix_i(t) \Big| Y^{0:t} \right]=\sum_{i=1}^{N}\gamma_i\hat{x}_i(t|t).
\end{align}
This policy is equivalent to select the $N_T$ individuals whose probability of being infected $\hat{x}_i(t|t)$ is the highest.

It should be noted that the proposed cost function is a primary attempt to define an efficient metric in line with the presented framework.
Nonetheless, the selection of the optimal cost function is out of the scope of this paper and remains an open problem for further research.

\section{Quarantine actions}\label{sec:quarantine}

The outcomes of the tests are exploited to act on the population through a selective quarantine. Formally, we introduce the control variable $q_k(t)$ such that $q_k(t)=1$ if $k$-th individual is selected to be quarantined at time instant $t$, and $q_k(t)=0$ otherwise. 
In this paper for any positive $i$ we propose to quarantine the $L$ closest neighbours, i.e. the $L$ individuals with the highest transmission probability $w_{ij}(t)$. The parameter $L$ can be properly tuned to trade-off between the total number of quarantined for positive and the expected number of infected (but not detected) that are quarantined because they have a direct contact with a positive. We consider that individuals will leave quarantine after $D_Q$ days. 

Note that in line of principle other quarantine strategies can be designed based on probabilities of infection of the neighbours of a positive tested, as well as preventive quarantine based only on the state estimate, and they will be the subject of future investigations.

\section{Numerical simulations}\label{sec:numerical_simulations}

This section shows, through numerical simulations, the effectiveness of the proposed solution by comparing it to current approaches.

\subsection{Setting}
The simulation setup considers a closed population of 10'000 individuals with the following parameters regarding the spread of the disease
\begin{itemize}
	\item $R_0=2$. This value is equivalent to a virus with high spreading, e.g. the Covid-19, when no social distance measures are adopted~\citep{giordano_modelling_2020, salath_covid19_2020}.
	\item $0.1\%$ of the population is initially infected.
	\item $20\%$ of new infected present symptoms of the disease before the recovery, in agreement with \cite{ing2020covid} and \cite{lavezzo2020suppression}.
	\item $0.5\%$ of the population can be tested at each day, corresponding to $N_T=50$. This value is similar to the percentage of daily tests in South Korea or in USA at December 2020, see \url{https://covidtracking.com/data}.
	\item The $L=5$ closest individuals of each individual with positive test are put in quarantine for $D_Q=14$ days. When in quarantine, all the transmission probability $w_{ij}$ are reduced to $1/100$ of their normal value.
\end{itemize}

The population distribution can be conveniently represented through a weighted undirected graph, where each node represents an individual, an edge between two nodes represents an interaction between two individuals, and the weight is set equal to the probability of transmission $w_{ij}$. The graph topology has been generated to emulate a small-world network. This kind of graphs are characterized by the presence of clusters, which are subgraphs that are (almost) complete, and of short paths connecting (almost) any pair of nodes. They have been introduced to capture the evidence of human connections and have been widely studied in the literature, see \cite{de1978contacts} and \cite{watts1998collective}.
In our case, each individual belongs to more clusters (at least 2, up to 6) to mimic families, offices, habitual relations and activities, etc. 
Dimensions of the clusters are uniformly distributed and the range depends on the kind of relationship that they capture: for example dimensions of households randomly vary from 1 to 8, while dimensions of offices vary from 4 to 40. Random links are also added to the network. The resulting graph is then heterogeneous and possibly unbalanced.
Average weights are set in a realistic way, e.g. average weights in a household are four times the ones in a small office. For the sake of simplicity the graph is assumed to be time-invariant, except for the effects of quarantine action.

Initial conditions $\xi(0)$, i.e. which individuals are initially infected, are stochastically generated based on the initial probability of each node to be infected. To test the robustness of the proposed strategy, we assume that the probability distribution of the initial conditions is perturbed up to the $10\%$. It is also assumed that $10\%$ of the arcs of the graph are unknown.

The presented simulations compare three different scenarios:
\begin{itemize}
	\item \textit{Test and trace (T\&T)}. This policy tracks the contacts (based on the knowledge of the network) of symptomatic and detected cases (see \cite{ferretti2020quantifying} and \cite{dar2020applicability}). More formally we define the set of individuals that have been detected at a generic time $\tau$ as $\mathcal{D}(\tau)= \{i\in\mathcal{S}(\tau) \,:\, x_i(\tau)=1\}$. Then for any individual $i\in\mathcal{D}(t\!-\!1)$ we retrieve the set of recent contacts $\mathcal{N}_i(t)=\{j \,:\, w_{ij}(\tau)>0,\,t\!-\!D\!<\!\tau<t\}$, and we refine it by removing already detected individuals and individuals that have been recently tested. From the set $\mathcal{N}_i(t)$ we select the individuals that have been more in contact with the individual $i\in\mathcal{D}(t\!-\!1)$. Among the different possibilities for doing so, we choose $\overline{\mathcal{N}}_i(t) \subseteq \mathcal{N}_i(t)$ such that 
		\begin{align}
		&w_{ik}(t\!-\!1)\geq w_{i\ell}(t\!-\!1) \  \ \forall k \in \overline{\mathcal{N}}_i(t) \ \forall \ell \in \mathcal{N}_i(t) \\
		&\left|\bigcup\nolimits_{i\in\mathcal{D}(t\!-\!1)} \overline{\mathcal{N}}_i(t)\right|=N_T \\
		&|\overline{\mathcal{N}}_i(t)| - |\overline{\mathcal{N}}_j(t)| \in \{-1,0,1\}
		\end{align}
		where $|\cdot|$ denotes the cardinality of the set.
		Less rigorously, we can say that $\overline{\mathcal{N}}_i(t)$ contains the contacts of $i$ with which the last interaction has been the most dangerous, while the number of individuals in each set $\overline{\mathcal{N}}_i(t)$ is chosen such that tests are allocated as uniformly as possible among the sets $\mathcal{N}_i(t)$ of contacts of detected positive. 
		Please note that, since in the following simulations the graph is fixed, $w_{ik}(t\!-\!1)\geq w_{i\ell}(t\!-\!1)$ implies that $w_{ik}(\tau)\geq w_{i\ell}(\tau)$ for any $\tau\leq t-1$ if $k,\ell$  has not been in quarantined.
		Remaining tests are used to randomly explore other parts of the graph in order to model test selection policies that are not based on the probabilities of infection, as it is done in reality where tests are also partially allocated to the employs of interested companies. Test and trace strategy is a well-known policy which has provided good results in several countries and it has been considered the best practice by the medical community \cite{who2020contact}.
	\item \textit{Topology-based testing}. This scenario presents a policy where, based on the topology of the graph, i.e. the number of contacts, certain individuals are periodically tested. In particular we choose a period $T=20\,$days and we solve the constrained optimization problem
	\begin{align}
		\gamma^T&=\arg\max_{\gamma}\ \sum\nolimits_i \sum\nolimits_j w_{ij}(0) \gamma_i\\
		&\text{s.t. } \sum\nolimits_i\gamma_i \leq TN_T
	\end{align}
	Individuals such that $\gamma^T_i=1$ are then randomly sorted and tested accordingly. The periodical testing campaign is on-line delayed in order to allocate tests to the closets neighbours of a new detected. However we consider only a partial tracing, assuming that at most 8 contacts are provided. The application of topology-based centrality metrics for test selection is quite new but they have been studied to select the edges to remove (see \cite{doostmohammadian2020centrality} and references therein).
	\item \textit{Smart testing (T\&EST)}. This scenario follows the proposed control scheme where individuals are selected according to the probability of being infected, eq.~\eqref{eq:cost}, and the outcomes of the tests are used to update the state estimate according to Sec. \ref{sec:approx_estimation}. 
	The correction term is set as
	\begin{equation}
	\sigma_{ij}(w_{ij}(t),w_{ij}(t\!-\!1), \dots, w_{ij}(0))) = (1-\epsilon)^{|\mathcal{T}_{ij}(t)|}
	\end{equation}
	with
	\begin{equation}
	\mathcal{T}_{ij}(t)=\{ \tau>0 : t\!-\!D \!<\! \tau \!<\! t, w_{ij}(\tau)>0 \}
	\end{equation}
	so $|\mathcal{T}_{ij}(t)|$ is the number of days in the last $D$ with a contact between individual $i$ and $j$. We set parameter $\epsilon$ by fitting the time evolution of the mean probability computed by the proposed estimator to the incidence of cases obtained by simulating a graph of a similar topology in the case of no actions on the system.
\end{itemize}

Given the stochastic nature of the model, $100$ simulations have been generated for each scenario. The spread of the infectious disease is monitored for a time span of $300$ days.

\subsection{Results}

\begin{figure}[b]
	\begin{minipage}{\columnwidth}
		\centering
		\includegraphics[width=0.9\columnwidth]{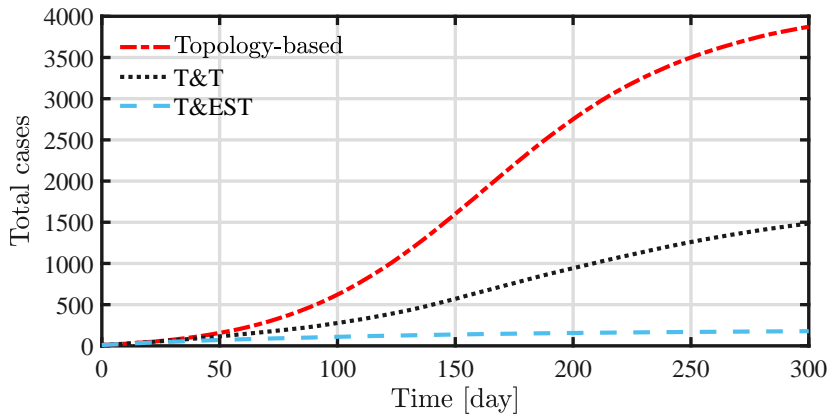}
		\vspace*{-10pt}
		\caption{Evolution of the cumulative number of infected individuals.\label{fig:cumulative}}
	\end{minipage}
\end{figure}

As we can see in Fig. \ref{fig:cumulative}, the proposed control mechanism (testing based on estimation and conditional quarantine) is effective in reducing the total number of infected people in a given temporal window both with respect to the topology-based strategy and test-and-trace. 
It is important to note that the test-and-trace strategy clearly outperforms the topology-based strategy even if the latter, when no new known positive are present, allocates tests on crucial points of the network instead of on random individuals. This result confirms how important is to trace contacts of positive individuals in order to keep the epidemic under control. 
On the other hand, the comparison between T\&T and T\&EST shows that the performance of pure tracing can be largely improved by using control methodologies. In particular, when many positive individuals have been detected, it is impossible to test all of their contacts due to the limited number of available tests. In that case, with the proposed strategy, the contacts that should be tested naturally come up among all the other contacts. When no new positive is detected, T\&EST selects who to test using updated information on the state of the population. Thus, the proposed strategy takes into account both information on the graph topology (for individuals that has not been in contact with tested, estimation is affected by the number of interactions) and the information from tests.  Obtained results show that including dynamics provides better performances than simpler off-line strategies.

\begin{figure}
	\begin{minipage}{\columnwidth}
		\centering
		\includegraphics[width=0.9\columnwidth]{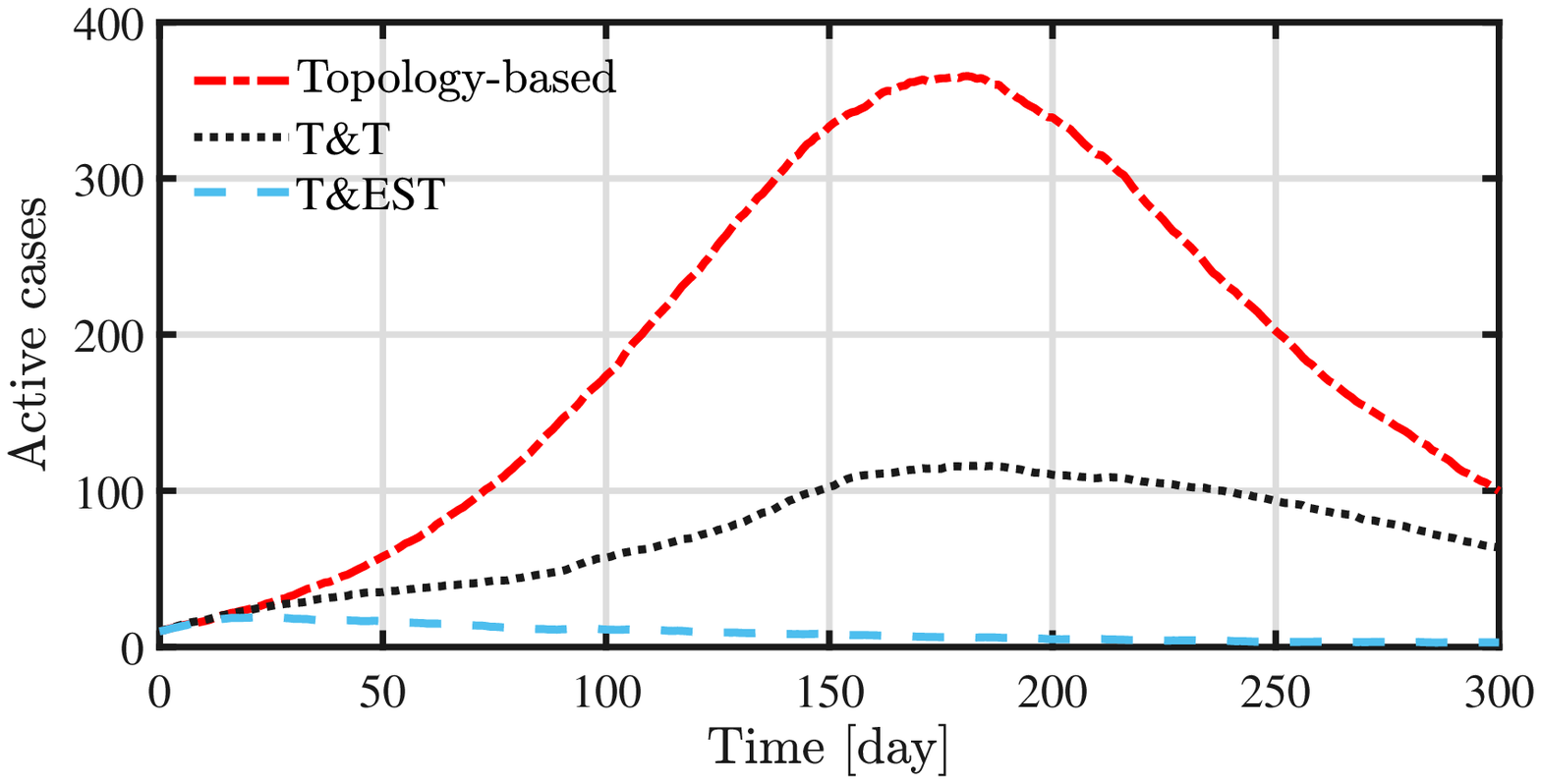}
		\vspace*{-10pt}
		\caption{Evolution of the  number active cases.\label{fig:current}}
		\vspace*{20pt}
	\end{minipage}
	\begin{minipage}{\columnwidth}
		\centering
		\includegraphics[width=0.9\columnwidth]{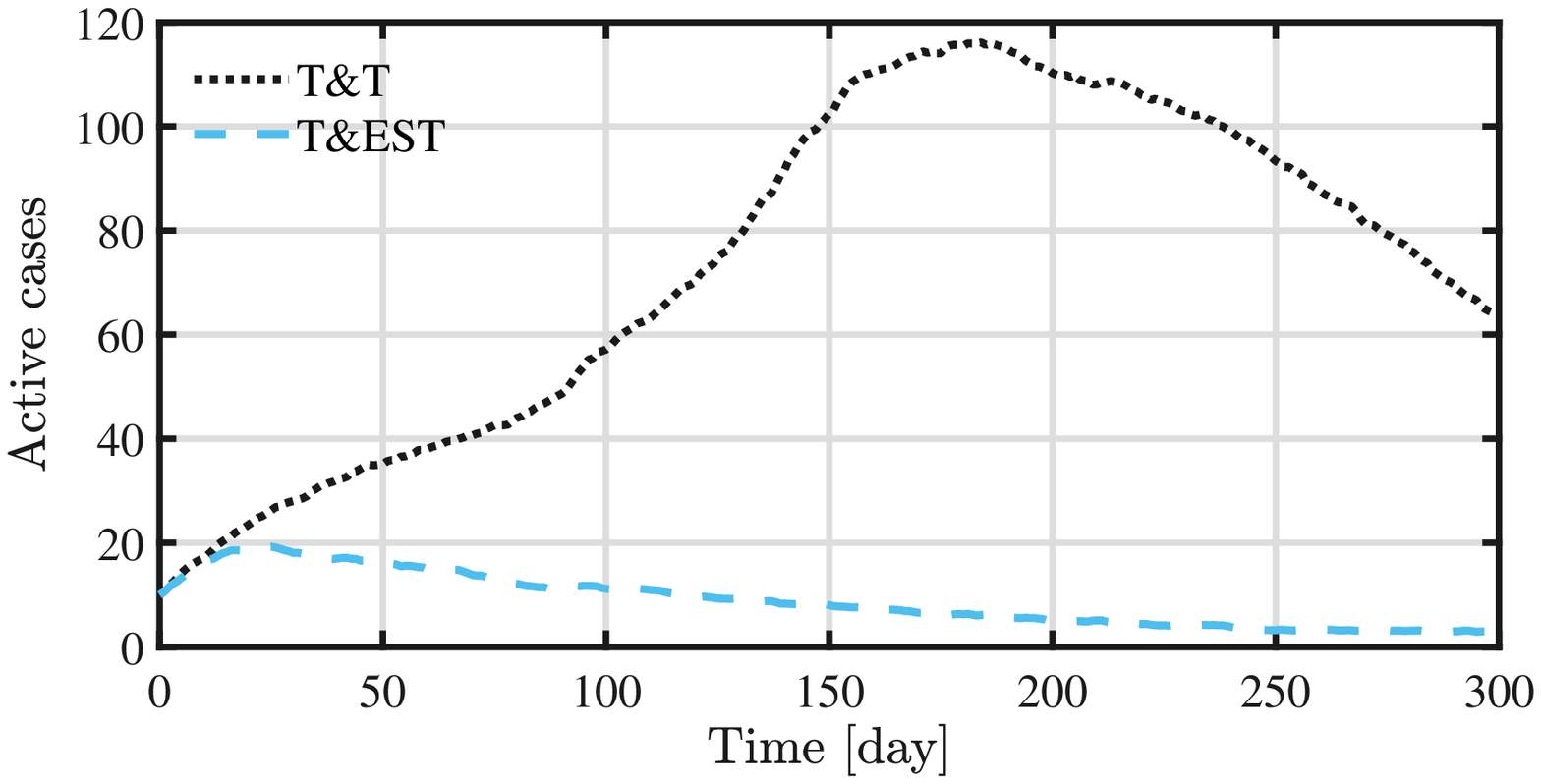}
		\vspace*{-10pt}
		\caption{Evolution of the  number active cases for test-and-trace and the presented strategy.\label{fig:current_zoom}}
	\end{minipage}
\end{figure}

The proposed strategy is effective also in mitigating the epidemic outbreak by avoiding any peak of active cases, as shown in Fig. \ref{fig:current}. This is an important result since it is fundamental to have a low number of active cases to avoid the health-care system maximum capacity to be reached. It is important to remark that the presented approach has shown to be very effective when applied to cases where the initial number of infected individuals is small compared to the total size of the population. In such a case, a fast identification of clusters of infection is essential, providing enough \textit{tracing} to detected positive individuals but also exploring new areas of the network. This improvement in the performance can be better appreciated in Fig~\ref{fig:current_zoom}.

\begin{figure}[ht!]
	\centering
	\includegraphics[width=0.9\columnwidth]{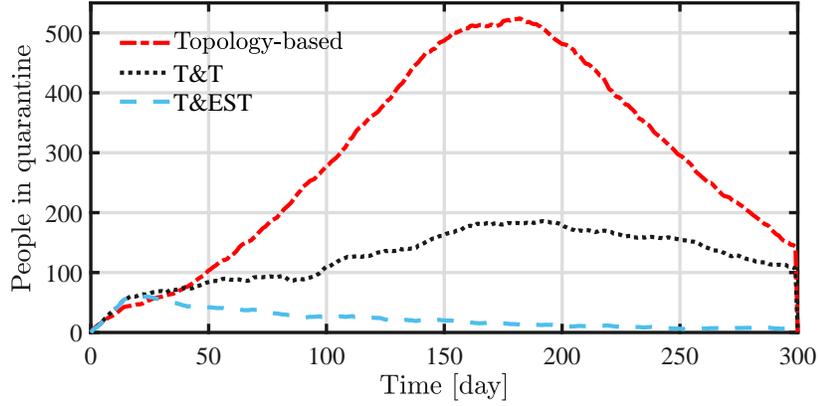}
	\vspace*{-10pt}
	\caption{Evolution of the  number of individuals in quarantine.\label{fig:quarantine}}
\end{figure}

The number of people in quarantine at each time instant is depicted in Fig.~\ref{fig:quarantine}. Although not intuitive, this plot shows lower numbers of people in quarantine for the smart testing policy, indicating that the improvement in performance does not require a greater number of people in quarantine but that actually can be achieved with less but better focused quarantines. In these simulations the number of people in quarantine for the T\&EST is almost negligible, showing that an efficient testing policy can have a great impact also in the required control actions. This is a very promising result especially from an economic point of view since it would limit the social and economical impact of the measures. 

A synoptic overview of the numerical simulations is reported in Table~\ref{tab:results_summary}. The results show the clear improvement on the containment of the epidemic, in terms of both active cases and people in quarantine, by using a testing and quarantine policy based on the presented  estimation algorithm.

\begin{table}[ht!]
	\centering
	\caption{Summary of the averaged results for the $3$ scenarios.}\label{tab:results_summary}
	\begin{tabular}{c c  c c} 
		\hline
		Scenario & Peak of active cases &  Total infected & Work days lost \\ [0.5ex] 
		\hline
		Test and trace & $116$ & $ 1472$ & $2564$  \\
		Topology-based strategy & $365$ & $3859$ & $5847$ \\
		Smart testing & $20$ & $169$ & $444$ \\
		\hline
	\end{tabular}
\end{table}

\begin{figure}[b]
	\centering
	\includegraphics[width=0.9\columnwidth]{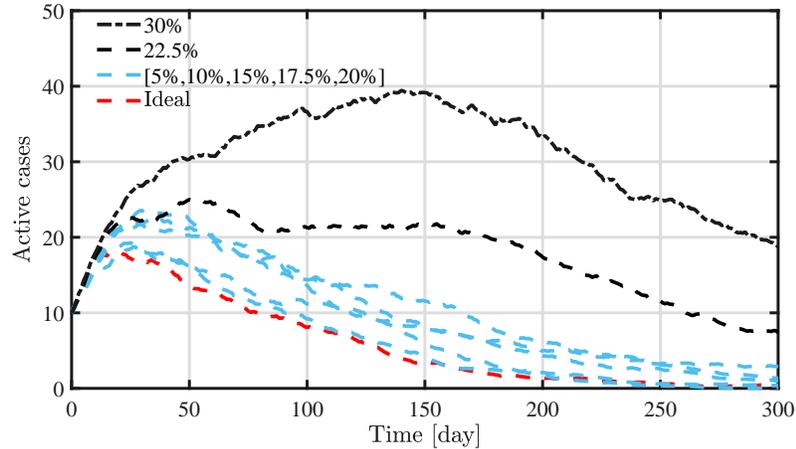}
	\vspace*{-20pt}
	\caption{Variation of the strategy performance based on the knowledge of the network.\label{fig:robustnes}}
\end{figure}

An important aspect of the presented strategy is the assumption of a good knowledge of the network topology. In this sense, Fig.~\ref{fig:robustnes} provides the variation in the performance of the presented strategy with respect to the number of individual interactions known. From this plot it can be seen that within the range of $80\%-100\%$ of knowledge of the network, the results are very similar and promising. For a percentage of unknown interactions superior to $20\%$, a threshold behaviour can be seen, where the performance is clearly worse and a more evident peak of infection can be seen.

\section{Conclusions}\label{sec:conclusions}

In this paper we presented a novel testing strategy to smartly select the individuals to be tested during an epidemic. This policy is based on a decentralized state estimation of the status of the epidemic obtained from the outcome of the tests.

The testing policy is defined as an optimization problem based on the state estimation. The proposed estimation algorithm is computationally inexpensive and can even be implemented in a distributed fashion.

The numerical results based on Montecarlo simulations demonstrate that the use of the proposed scheme, testing and selective quarantine, significantly reduces the total number of infected people as well as the peak of active case and the number of people put in quarantine. 

Future works will focus on the link between the test selection objective functions and the quarantine policies. The case where the reliability of the test is considered is another subject of study in future research.

\bibliographystyle{plainnat}
\bibliography{main}

\end{document}